\documentclass[10pt,aps,prb,twocolumn,showpacs,amssymb,floatfix]{revtex4-1}
\usepackage{amsmath,graphics,epsfig,mathrsfs}
\usepackage{epstopdf}
\usepackage{bm}
\usepackage{color}
\usepackage{soul}
\usepackage{hyperref}
\hypersetup{backref=true,pagebackref=true,hyperindex=true,colorlinks=true,breaklinks=true,citecolor=blue,urlcolor=blue,linkcolor=blue,bookmarks=true,bookmarksopen=true}

\begin{document}
\title{Spin orbit torque in disordered antiferromagnets}
\author{Hamed Ben Mohamed Saidaoui}
\affiliation{Physical Science and Engineering Division, King Abdullah University of Science and Technology (KAUST), Thuwal 23955-6900, Kingdom of Saudi Arabia}
\author{Aurelien Manchon}
\email{aurelien.manchon@kaust.edu.sa}
\affiliation{Physical Science and Engineering Division, King Abdullah University of Science and Technology (KAUST), Thuwal 23955-6900, Kingdom of Saudi Arabia}

\begin{abstract}
We investigate the current-induced spin-orbit torque in antiferromangnetic materials in the presence of Rashba spin orbit coupling using both the linear response theory and the non-equilibrium Green's function technique implemented on a tight-binding model. We show that a staggered spin density arises from interband contributions. The effect of the disorder is restricted to the diminution of the torque magnitude and does not bring any anomalous behavior to the dependences of the torque on the spin orbit strength and the exchange interaction between the antiferromagnetic local moments and the itinerant electrons spins. We prove that the spin orbit torque is robust against the application of the disorder compared to the case of the antiferromagnetic spin-valve spin transfer torque.
\end{abstract}
\maketitle

\section{Introduction}

Twenty years ago, Slonczewski \cite{slonczewski1996} and Berger \cite{berger1996} predicted simultaneously and independently that a non-equilibrium spin current injected in a ferromagnet exerts a torque on the local magnetization, thereby enabling its efficient electrical control. This mechanism, referred to as spin transfer torque, has been intensively investigated in metallic and tunneling spin-valves \cite{Ralph2008,Brataas2012,Tsoi2011,Manchon2011}, i.e. devices comprising a reference magnetic layer playing the role of a spin-polarizer and a free magnetic layer that absorbs the incoming spin current. A crucial feature of spin transfer torque is the need for non-collinearity between the reference and free layers' magnetization directions. This concept has been soon extended to magnetic domain walls \cite{Tserkovnyak2008,Boulle2011}, which have attracted a massive amount of interest lately due to their gigantic current-driven velocities (see, e.g. Refs. \onlinecite{Moore2008,Miron2011,Yang2015}). \par

Ten years later N\'u\~nez et al. \cite{nunez2006,duine2007} extended the concept of spin torque to antiferromagnets. The authors computed the torque present in a spin-valve composed of two antiferromagnetic chains separated by a metallic spacer and demonstrated that the staggered spin density generated in the first antiferromagnet can be "imprinted" on the second antiferromagnet thereby exerting a torque, without net spin transfer \cite{Haney2008}. This prediction drew a significant amount of excitement as antiferromagnets possess a number of properties of interest for applications, such as the absence of demagnetizing field and most importantly, the possibility for ultrafast (TeraHertz) manipulation of the order parameter as emphasized recently \cite{Kimel2004,Kimel2009,Wienholdt,Cheng2015}. Although a number of theoretical studies \cite{xu2008,haney2007} have confirmed the general picture proposed by N\'u\~nez et al., the experimental evidence of such a torque has remained particularly challenging \cite{urazhdin2007,wei2007}, as this torque is extremely sensitive to disorder. In fact, since the spin torque in antiferromagnetic spin-valves involve the {\em coherent transmission} of a staggered spin density from one antiferromagnet to another, it is dramatically altered by the presence of disorder-driven momentum scattering \cite{saidaoui2014,duine2007}. To overcome this limitation, one needs to generate a torque that is immune from momentum scattering. The first option is to use a tunnel barrier to separate the two antiferromagnetic electrodes. In this case, the torque is controlled by the interfacial density of states and is therefore less sensitive to the disorder inside the antiferromagnetic layers \cite{merodio,saidaoui2016}. Another option is to generate a {\em local} torque, i.e. a torque that does not need any transfer of spin information from one part of the device to the other. One way is to use antiferromagnetic domain walls where the torque is generated locally by the domain texture \cite{Hals2011,Tveten2013}. Another way is to use spin-orbit interaction in antiferromagnetic systems lacking inversion symmetry \cite{zelezny2014}. \par

As a matter of fact, several authors \cite{bernevig_2005,manchon2008,garate2009} recently suggested that in magnetic materials presenting both (bulk or interfacial) inversion symmetry breaking and strong spin-orbit coupling, a flowing charge current generates a non-equilibrium spin density (a phenomenon known as inverse spin galvanic \cite{Ivchenko} or Edelstein effect \cite{edelstein1990}) that can in turn directly exert a torque on the local magnetization, without the need for an external polarizer. This mechanism, called spin-orbit torque, has attracted intense attention in the past decade, with the possibility to use spin-orbit torques in three terminal devices for switching \cite{Miron2011b,Liu2012,Fan2014} and excitations \cite{Liu2012b,demidov2012}. Among the recent theoretical progress, the importance of interband contributions to the spin-orbit torque has been emphasized by Kurebayashi et al. \cite{Kurebayashi2014} and investigated in details in several theoretical studies \cite{bijl2012,Li2015,Lee2015}.\par

Spin-orbit coupled effects have been lately recognized to play a central role in antiferromagnets (see e.g., Ref. \onlinecite{Jungwirth2016}). A large anisotropic tunneling magnetoresistance was observed in IrMn-based tunnel junctions \cite{park2011,Marti2012,Wang2012,Wang2013}, followed by demonstrations of anisotropic magnetoresistance in FeRh \cite{Marti2013}, and Sr$_2$IrO$_4$\cite{Wang2014}. In a recent work, \v Zelezn\'y et al. \cite{zelezny2014,zelezny2016} predicted that spin-orbit torques can be used to excite the order parameter of collinear bipartite antiferromagnets. Indeed, this study demonstrated that in a Rashba two-dimensional electron gas with antiferromagnetic order, a "N\'eel" torque on the form ${\bf T}\sim{\bf n}\times({\bf p}\times{\bf n})$ emerges, where ${\bf n}$ is the order parameter and ${\bf p}$ is a unit vector determined by the crystal symmetries. In other words, this torque acts like a {\em staggered} magnetic field that changes sign on the two sublattices of the antiferromagnet, therefore inducing a coherent precession of the N\'eel order parameter. Spin-orbit torque were also calculated in Mn$_2$Au, an antiferromagnet with a {\em hidden} symmetry breaking \cite{barthem2013}. In this system, while the non-magnetic crystal structure is inversion symmetric, each magnetic sublattice possesses an opposite symmetry broken environment. In this second case, the torque has the form ${\bf T}\sim{\bf n}\times{\bf p}$, also acting like a staggered magnetic field. The experimental demonstration of the spin-orbit torque in bulk antiferromagnets was reported very recently by Wadley et al. \cite{wadley2015} in CuMnAs, an antiferromagnet also presenting hidden bulk inversion symmetry breaking \cite{maca2012,wadley2013}. Preliminary evidence of interfacial spin-orbit torques in Ta/IrMn has been reported recently \cite{Reichlova}.\par

The objective of the present work is to address the nature of the spin-orbit torque in antiferromagnetic Rashba two-dimensional electron gas using both analytical calculations and real-space tight-binding model. In particular, we are interested in evaluating the robustness of this torque against disorder. In Section \ref{s:kubo}, a free electron model for antiferromagnetic Rashba two-dimensional electron gases is exposed and the spin-orbit torque is computed using Kubo formula in the weak disorder limit. In Section \ref{s:tb}, a tight-binding model is introduced and a detailed analysis of the torques is presented, with a particular attention to the role of disorder in Section \ref{s:dis}. Section \ref{s:conclusion} presents the conclusions of this study.

 \section{Analytical Derivation of Spin-Orbit Torques\label{s:kubo}}
This section presents the calculation of the non-equilibrium spin density in an antiferromagnetic two dimensional electron gas with Rashba spin-orbit coupling. While these results have been briefly summarized in Ref. \onlinecite{zelezny2016}, we hereby provide the detailed derivation and an in-depth discussion of the model. This analysis will serve as a guideline to examine the numerical results of Section \ref{s:tb}.

\subsection{System definition}
We first derive the effective low energy Hamiltonian for a two-dimensional antiferromagnet constituted of two antiferromagnetically coupled sublattices on a square lattice, with Rashba spin-orbit interaction. The antiferromagnetic configuration, illustrated in Fig. \ref{fig:introductory}, is G-type or checkerboard. Namely each magnetic moment is surrounded by antiferromagnetically coupled nearest neighbor moments. The Bloch wave function of the crystal is given by
\begin{eqnarray}\label{eq:1}
\Psi_{\bf k}({\bf r})&=&\frac{1}{\sqrt{N}}\sum_{j}e^{i{\bf k}\cdot({\bf r}-{\bf R}_j)}\left[\hat{\varphi}_{\rm A}({\bf r}-{\bf R}_j)\right.\\
&&\left.+e^{-i{\bf k}\cdot{\bf R}_{\rm B}}\hat{\varphi}_{\rm B}({\bf r}-{\bf R}_{\rm B}-{\bf R}_j)\right],\nonumber
\end{eqnarray}
where $\hat{\varphi}_i({\bm r})$ are the spin-polarized states on sublattice sites $i=$A, B. The magnetic moments of A and B are aligned along $\pm{\bf n}$, ${\bf n}$ being the N\'eel vector of the system. The summation runs over all the $N$ unit cells of the system. In our notation, ${\bf R}_j$ is the position of the $j$-th unit cell and ${\bf R}_{\rm B}$ is the relative position of atom B with respect to atom A in the unit cell. Our objective is to express the Hamiltonian, Eq. \eqref{eq:1}, in the basis $\{|\text{A}\rangle,|\text{B}\rangle\}\otimes\{|\uparrow\rangle,|\downarrow\rangle\}$, where $|i\rangle$ refers to the states on sublattice $i$ and $|\sigma\rangle$ refers to the local spin projection on ${\bf n}$. The energy of the system is therefore given by
\begin{eqnarray}\label{eq:3}
E({\bf k})=\int \frac{d{\bf r}}{\Omega}\Psi_{\bf k}^\dagger({\bf r}){\hat H}\Psi_{\bf k}({\bf r})
\end{eqnarray}
where ${\hat H}$ is the real-space tight-binding Hamiltonian of the two-dimensional crystal given in Eq. (\ref{eq:htb}) (see also Ref. \onlinecite{zelezny2014}). The effective Hamiltonian in k-space is obtained by inserting Eq. (\ref{eq:1}) into Eq. (\ref{eq:3}), and considering only nearest neighbor hopping, which leads to
\begin{eqnarray}\label{eq:hf}
\tilde{H}=&&-4t\cos(\tilde{k}_x-\tilde{k}_y)\cos(\tilde{k}_x+\tilde{k}_y)\hat\tau_x\\
&&-(\alpha/a)(\sin\tilde{k}_y\hat\sigma_x-\sin\tilde{k}_x\hat\sigma_y)\hat\tau_x+J_{\rm sd}{\bf n}\cdot\hat{\bm\sigma}\hat\tau_z,\nonumber
\end{eqnarray}
where $t$ is the hopping energy between nearest neighbors, $\alpha$ is the Rashba parameter (in eV.m), $J_{\rm sd}$ is the exchange energy between the local moments and itinerant spin, $\tilde{k}=ka/2$ and $a$ is the lattice parameter. Notice that $\alpha/2a=t_{so}$ in the notation of Eq. (\ref{eq:htb}). The Pauli operators for spin 1/2, $\hat{\bm\sigma}$ and $\hat{\bm\tau}$, apply to the subspace of real spin $\{|\uparrow\rangle,|\downarrow\rangle\}$ and to the subspace of sublattice sites $\{|\rm A\rangle,|\rm B\rangle\}$, respectively. To the second order in $k$ in the limit $k\rightarrow 0$, i.e. close to the $\Gamma$ point, Eq. (\ref{eq:hf}) reduces to
\begin{eqnarray}\label{eq:hfk}
\tilde{H}=&\gamma_k\hat\tau_x-\alpha k\hat{\bm\sigma}\cdot{\bm\mu}\hat\tau_x+J_{\rm sd}{\bf n}\cdot\hat{\bm\sigma}\hat\tau_z,
\end{eqnarray}
where $\gamma_k=ta^2\left(k^2-k_0^2\right)$, ${\bm \mu}=(\sin\varphi_k,-\cos\varphi_k,0)$, $k_0=2/a$ and ${\bf k}=(\cos\varphi_k,\sin\varphi_k,0)$. The latter, Eq. \eqref{eq:hfk}, will be the basis of the analytical derivation of the spin-orbit torque.

The associated {\em unperturbed} retarded Green's function, defined as $\hat{G}^R_0=(\epsilon-\hat{H}+i0^+)^{-1}$, reads
\begin{eqnarray}\label{eq:go}
\hat{G}^R_{0}&=&\frac{1}{4\zeta_k}\sum_{s,\eta=\pm1}\frac{1}{\epsilon-\epsilon_{{\bf k},s,\eta}+i0^+}\times\\
&&\left[\zeta_k+s(\gamma_k(\hat{\bm\sigma}\cdot{\bm\mu})-J_{\rm sd}\hat{\tau}_y(\hat{\bm\sigma}\cdot{\bf n}\times{\bm\mu}))\right.\nonumber\\
&&\left.+\frac{1}{\epsilon_{{\bf k},s,\eta}}\left((s\gamma_k^2+sJ_{\rm sd}^2+\alpha k \zeta_k)(\hat{\bm\sigma}\cdot{\bm\mu})\hat{\tau}_x\right.\right.\nonumber\\
&&\left.\left.+(\gamma_k\hat{\tau}_x+J_{\rm sd}(\hat{\bm\sigma}\cdot{\bf n})\hat{\tau}_z)(\zeta_k+s\alpha k)\right.\right.\nonumber\\
&&\left.\left.-sJ_{\rm sd}({\bf n}\cdot{\bm\mu})(J_{\rm sd}(\hat{\bm\sigma}\cdot{\bf n})\hat{\tau}_x-\gamma_k\hat{\tau}_z+\alpha k(\hat{\bm\sigma}\cdot{\bm\mu})\hat{\tau}_z)\right)\right]\nonumber
\end{eqnarray}
where 
\begin{eqnarray}
\epsilon_{{\bf k},s,\eta}&=&\eta\sqrt{\gamma_k^2+J_{\rm sd}^2+\alpha^2k^2+2s\alpha k\zeta_k},\\
\zeta_k&=&\sqrt{\gamma_k^2+J_{\rm sd}^2(1-\sin^2\theta\sin^2(\varphi_k-\varphi))}.\label{eq:sk}
\end{eqnarray}

In order to get an analytically tractable expression, in the following we express the Green's function in term of the projection operator ${\cal A}_{s,\eta}=|s,\eta\rangle\langle s,\eta|$ such that $\hat{G}^{R}_{0}=\sum_{s,\eta}{\cal A}_{s,\eta}/(\epsilon-\epsilon_{{\bf k},s,\eta}+ i0^+)$. 

\subsection{Kubo Formalism}

To compute the non-equilibrium spin density, we now use the Kubo formula by assuming that the presence of {\em weak} spin-independent impurities only broadens the energy levels by an amount $\Gamma$. Hence, the perturbed Green's functions read $\hat{G}^{R,A}=1/(\epsilon-\hat{H}\pm i\Gamma)=\sum_{s,\eta}{\cal A}_{s,\eta}/(\epsilon-\epsilon_{{\bf k},s,\eta}\pm i\Gamma)$. In the limit of weak disorder, $\Gamma\rightarrow0$, the non-equilibrium spin density driven by an electric field $e{\bf E}$ possesses two main contributions (see for instance Ref. \onlinecite{Li2015})
\begin{eqnarray}\label{eq:intra}
{\bf s}^{\rm Intra}=&&\frac{e\hbar}{2\Gamma\Omega}\sum_{\nu,{\bf k}}{\rm Re}\{{\rm Tr}[(\hat{\bm v}\cdot{\bf E}){\cal A}_\nu\hat{\bm\sigma}_\varsigma{\cal A}_\nu]\}\delta(\epsilon_{{\bf k},\nu}-\epsilon_{\rm F}),\\
{\bf s}^{\rm Inter}=&&\frac{e\hbar}{\Omega}\sum_{\nu\neq \nu',{\bf k}}{\rm Im}\{{\rm Tr}[(\hat{\bm v}\cdot{\bf E}){\cal A}_\nu\hat{\bm\sigma}_\varsigma{\cal A}_{\nu'}]\}\frac{(f_{{\bf k},\nu}-f_{{\bf k},\nu'})}{(\epsilon_{{\bf k},\nu}-\epsilon_{{\bf k},\nu'})^2}\nonumber\\\label{eq:inter2}\end{eqnarray}
where $\nu=s,\eta$ for conciseness. Here, $\hat{\bm v}$ is the velocity operator, $f_{{\bf k},\nu}$ is the Fermi-Dirac distribution of state ${\bf k},\nu$, $\epsilon_{\rm F}$ is the Fermi energy, and $\Omega$ is the volume of the Brillouin zone. We also define $\hat{\bm\sigma}_\varsigma=\hat{\bm\sigma}(1+\varsigma\hat{\tau}_z)/2$ ($\varsigma=+1$ for A and -1 for B sublattices), which defines the spin density operator on the A and B sublattices. \par

Since evaluating the transport properties involves an angular averaging over $\varphi_k$, it is convenient to rewrite the projection operator in the form ${\cal A}_{s,\eta}={\cal A}_{s,\eta}^{\rm e}+{\cal A}_{s,\eta}^{\rm o}$, where the first term is even in ${\bf k}$, while the second term is odd in ${\bf k}$. Furthermore, from now on we will only focus on the limit case of $\theta\ll 1$ (i.e. ${\bf n}\approx{\bf z}$) in order to get rid of the angular dependence of the Fermi surface contained in $\zeta_k$, Eq. (\ref{eq:sk}). This way, the energy dispersion reduces to $\epsilon_{s,\eta}=\eta(\sqrt{\gamma_k^2+J_{\rm sd}^2}+s\alpha k)$ and we find explicitly
\begin{eqnarray}\label{eq:aeven}
&&{\cal A}_{s,\eta}^{{\rm e}}=(1/4)\left[1+\eta\cos\theta_k\hat{\tau}_x+\eta\sin\theta_k(\hat{\bm{\sigma}}\cdot{\bf n})\hat{\tau}_z\right],\\\label{eq:aodd}
&&{\cal A}_{s,\eta}^{{\rm o}}=(s/4)\hat{\bm{\sigma}}\cdot\left[\cos\theta_k{\bm\mu}-\sin\theta_k\hat{\tau}_y({\bf n}\times{\bm\mu})+\eta{\bm\mu}\hat{\tau}_x\right],
\end{eqnarray}
where we defined $\cos\theta_k=\gamma_k/\sqrt{\gamma_k^2+J_{\rm sd}^2}$ and $\sin\theta_k=J_{\rm sd}/\sqrt{\gamma_k^2+J_{\rm sd}^2}$. Using the definitions of Eqs. (\ref{eq:intra})-(\ref{eq:inter2}) we notice that the heart of the physics is contained in the trace
\begin{equation}\label{eq:trace}
{\rm Tr}[(\hat{\bm v}\cdot{\bf E}){\cal A}_\nu\hat{\bm\sigma}_\varsigma{\cal A}_{\nu'}]=\langle\nu|\hat{\bm\sigma}_\varsigma|\nu'\rangle\langle\nu'|(\hat{\bm v}\cdot{\bf E})|\nu\rangle
\end{equation}
The velocity operator $\hat{\bm v}\cdot{\bf E}=\partial_{\bf k}\tilde{H}\cdot{\bf E}/\hbar$ close to the $\Gamma$-point reads
\begin{eqnarray}
\hbar\hat{\bm v}\cdot{\bf E}=&(2ta^2 {\bf k}\cdot{\bf E}+\alpha({\bf z}\times{\bf E})\cdot\hat{\bm \sigma})\hat{\tau}_x.
\end{eqnarray}

The trace Eq. (\ref{eq:trace}) has then two contributions that do not vanish up $\varphi_k$-integration,
\begin{eqnarray}\label{eq:trace2}
{\rm Tr}_{\rm d}^{\nu,\nu'}&=&\frac{2ta^2}{\hbar}({\bf k}\cdot{\bf E}){\rm Tr}[\hat{\tau}_x({\cal A}^{\rm o}_\nu\hat{\bm\sigma}_\varsigma{\cal A}^{\rm e}_{\nu'}+{\cal A}^{\rm e}_\nu\hat{\bm\sigma}_\varsigma{\cal A}^{\rm o}_{\nu'})],\\
{\rm Tr}_{\rm a}^{\nu,\nu'}&=&\frac{\alpha}{\hbar}{\rm Tr}\left[({\bf z}\times{\bf E})\cdot\hat{\bm\sigma}\hat{\tau}_x({\cal A}^{\rm o}_\nu\hat{\bm\sigma}_\varsigma{\cal A}^{\rm o}_{\nu'}+{\cal A}^{\rm e}_\nu\hat{\bm\sigma}_\varsigma{\cal A}^{\rm e}_{\nu'})\right].\nonumber\\\label{eq:trace3}
\end{eqnarray}

%

After some algebra, we obtain the following expression for the real and imaginary parts of the trace defined in Eq. (\ref{eq:trace}),
\begin{eqnarray}\label{eq:traceRef}
{\rm Re}{\rm Tr}^{s,\eta}&=&\frac{\eta}{2\hbar}[2sta^2\cos\theta_k+\alpha/k]\cos\theta_k({\bf k}\cdot{\bf E}){\bm\mu},\\\label{eq:traceImf}
{\rm Im}{\rm Tr}^{s,\eta}&=&-s\varsigma \frac{ta^2}{\hbar}\sin\theta_k({\bf k}\cdot{\bf E})({\bf n}\times{\bm \mu})\delta_{s'+s},\delta_{\eta'+\eta},\nonumber\\
\end{eqnarray}
where Eq. (\ref{eq:traceRef}) involves only intraband transitions ($s,\eta=s',\eta'$), while Eq. (\ref{eq:traceImf}) involves only interband transitions ($-s,-\eta= s',\eta'$). We can now proceed with the k-integration.


\subsection{Analytical expressions}
Since the energy $\epsilon_{{\bf k},s,\eta}$ is isotropic (independent on $\varphi_k$), Eqs. (\ref{eq:traceRef}) and (\ref{eq:traceImf}) can be further simplified by performing the angular integration
\begin{eqnarray}
\int d\varphi_k{\rm Re}{\rm Tr}^{s,\eta}&=&\frac{\eta\pi}{2\hbar}[2sta^2k\cos\theta_k+\alpha]\cos\theta_k({\bf z}\times{\bf E}),\nonumber\\\label{eq:traceReint}\\
\int d\varphi_k{\rm Im}{\rm Tr}^{s,\eta}&=&-s\pi\varsigma \frac{ta^2k}{\hbar}\sin\theta_k({\bf n}\times({\bf z}\times{\bf E}))\delta_{s'+s},\delta_{\eta'+\eta}.\nonumber\\\label{eq:traceImint}
\end{eqnarray}
Using Eqs. (\ref{eq:intra}) and (\ref{eq:inter2}) and noticing that $\delta(\epsilon_{{\bf k},s,\eta}-\epsilon_{\rm F})=|2t_{\rm N}a^2k\cos\theta_k+s\alpha|^{-1}\delta_{k-k_{\rm F}^{s}}$ (where $k_{\rm F}^{s}$ is the solution of $\epsilon_{{\bf k},s,\eta}=\epsilon_{\rm F}$), we obtain
\begin{eqnarray}\label{eq:intra1}
{\bf s}^{\rm Intra}&=&\frac{{\bf z}\times e{\bf E}}{16\pi\Gamma}\int dk k\cos\theta_k(\delta_{k-k_{\rm F}^+}-\delta_{k-k_{\rm F}^-}),\\
{\bf s}^{\rm Inter}&=&-\varsigma\frac{J_{\rm sd} ta^2}{8\pi}({\bf n}\times({\bf z}\times e{\bf E}))\int_{k_{\rm F}^-}^{k_{\rm F}^+} \frac{k^2dk}{(\gamma_k^2+J_{\rm sd}^2)^{3/2}}\label{eq:inter21}\end{eqnarray}
We consider the Fermi energy close to the top of the upper bands, so that the lower bands remain fully occupied $f_{{\bf k}, s,-}=1$. Furthermore, we recognize that to the linear order in $\alpha$
\begin{eqnarray}
k_{\rm F}^s&\approx& k_{\rm F}^0+\alpha k^\alpha_{\rm F}+{\cal O}(\alpha^2)\\
&=&\frac{1}{\sqrt{ta^2}}\sqrt{4t-\sqrt{\epsilon_{\rm F}^2-J_{\rm sd}^2}}+s\frac{\alpha}{2ta^2}\frac{\epsilon_{\rm F}}{\sqrt{\epsilon_{\rm F}^2-J_{\rm sd}^2}}.
\end{eqnarray}
Then, the integral in Eq. (\ref{eq:inter21}) can be rewritten
\begin{eqnarray}
\int_{k_{\rm F}^-}^{k_{\rm F}^+} \frac{k^2dk}{(\gamma_k^2+J_{\rm sd}^2)^{3/2}}\approx 2\alpha k_{\rm F}^\alpha\frac{k_{\rm F}^{02}}{(\gamma_{k_{\rm F}^{0}}^2+J_{\rm sd}^2)^{3/2}}
\end{eqnarray}
Finally, we find that the current-driven spin densities read
\begin{eqnarray}
{\bf s}^{\rm Intra}_{\rm AF}=&&\frac{m^*\alpha}{8\pi\hbar^2\Gamma}\left(1+2\frac{J_{\rm sd}^2}{\epsilon_{\rm F}^2}\left[2-\frac{4t}{\sqrt{\epsilon_{\rm F}^2-J_{\rm sd}^2}}\right]\right){\bf z}\times e{\bf E},\nonumber\\\label{eq: Sintaraf}\\
{\bf s}^{\rm Inter}_{\rm AF}=&&-\varsigma\frac{m^*\alpha J_{\rm sd}}{4\pi \hbar^2\epsilon_{\rm F}^2}\left(1-\frac{4t}{\sqrt{\epsilon_{\rm F}^2-J_{\rm sd}^2}}\right){\bf n}\times({\bf z}\times{\bf E}),\label{eq: Sinteraf}
\end{eqnarray}
where we replaced $ta^2=\hbar^2/2m^*$.\par

It is clear that the intraband transitions produce a spin density $\sim{\bf z}\times e{\bf E}$ that does not depend on the magnetization direction, consistently with the theoretical results obtained for ferromagnetic Rashba systems \cite{manchon2008,bijl2012,Pesin2012,wangmanchon2012,Li2015}. The only influence of the exchange in this expression is related to the modification of the density of state. The contribution of the interband transition is more interesting since it produces a spin density that depends on the direction of the {\em local} magnetic moment direction (i.e. $\sim \varsigma[{\bf n}\times({\bf z}\times{\bf E})]$) and therefore is of opposite sign on the two sublattices. Therefore, this spin density is {\em staggered} and expected to induce coherent manipulation of the antiferromagnetic order parameter ${\bf n}$ as explained in Ref. \onlinecite{zelezny2014}. These analytical results thereby confirm the numerical results obtained in Ref. \onlinecite{zelezny2014}. As a last comment, we can compare these torques to the ones obtained using the very same method on a ferromagnetic Rashba gas \cite{Li2015}. In the strong exchange limit, the authors found
\begin{eqnarray}\label{eq: Sintraf}
{\bf s}^{\rm Intra}_{\rm F}=&&\frac{1}{2\pi}\frac{m\alpha}{\hbar^2\Gamma}{\bf z}\times e{\bf E},\\\label{eq: Sinterf}
{\bf s}^{\rm Inter}_{\rm F}=&&-\frac{1}{2\pi}\frac{m\alpha}{\hbar^2J_{\rm sd}}{\bf m}\times({\bf z}\times{\bf E}),
\end{eqnarray}
where ${\bf m}$ is the magnetization direction. While the intraband contributions, Eqs. (\ref{eq: Sintaraf}) and (\ref{eq: Sintraf}), are essentially the same in both systems \cite{comment}, the interband contributions, Eqs. (\ref{eq: Sinteraf}) and (\ref{eq: Sinteraf}), differ noticeably. In ferromagnets, $S^{\rm Inter}_{\rm F}\sim 1/J_{\rm sd}$, while in antiferromagnets, $S^{\rm Inter}_{\rm AF}\sim J_{\rm sd}/\epsilon_F^2$. This can be understood by the fact that the energy splitting between different bands $\epsilon_{{\bf k},\nu}-\epsilon_{{\bf k},\nu'}$ in strong ferromagnets are mostly proportional to $J_{\rm ex}$, while in the case of antiferromagnets, it is driven by Rashba spin-orbit coupling.\par

Before moving on, let us conclude this section by emphasizing the limits of applicability of the above formulae. First the Hamiltonian, Eq. \eqref{eq:hfk}, is obtained within the free electron approximation up to the second order in $k$. This is a strong assumption since in antiferromagnets, this approximation only holds very close to the bottom (top) of the lower (upper) band and quartic terms $\sim k^4$ rapidly become important away from $\Gamma$-point. Second, these results are obtained to the linear order of Rashba spin-orbit coupling, which means that $\alpha k_{\rm F}\ll \sqrt{\gamma_{k_{\rm F}}^2+J_{\rm sd}^2}$. Third, the formulae are derived when the magnetic order is close to the normal to the plane ${\bf n}\approx{\bf z}$, which indicates that one can reasonably expect strong angular dependence of the torque magnitude. Third, we neglected the vertex corrections. This assumption is quite strong since it has been clearly demonstrated that in a ferromagnetic Rashba gas, such corrections cancel the interband contribution to spin-orbit torque in the absence of spin-dependent momentum scattering \cite{Qaiumzadeh2015}. Finally, these results are obtained within the limit of weak impurities assuming an isotropic relaxation time approximation and does not address the crucial issue of the robustness of the torque against disorder.

\section{Tight-binding model\label{s:tb}}

Let us now turn our attention towards the numerical calculation of these spin-orbit torques in the presence of random disorder. To do so, we use a real-space tight-binding model of the antiferromagnet that provides further insight in terms of materials parameter dependence and robustness against disorder.

\subsection{System definition}

\begin{figure}[h!]
  \includegraphics[width=8.5cm]{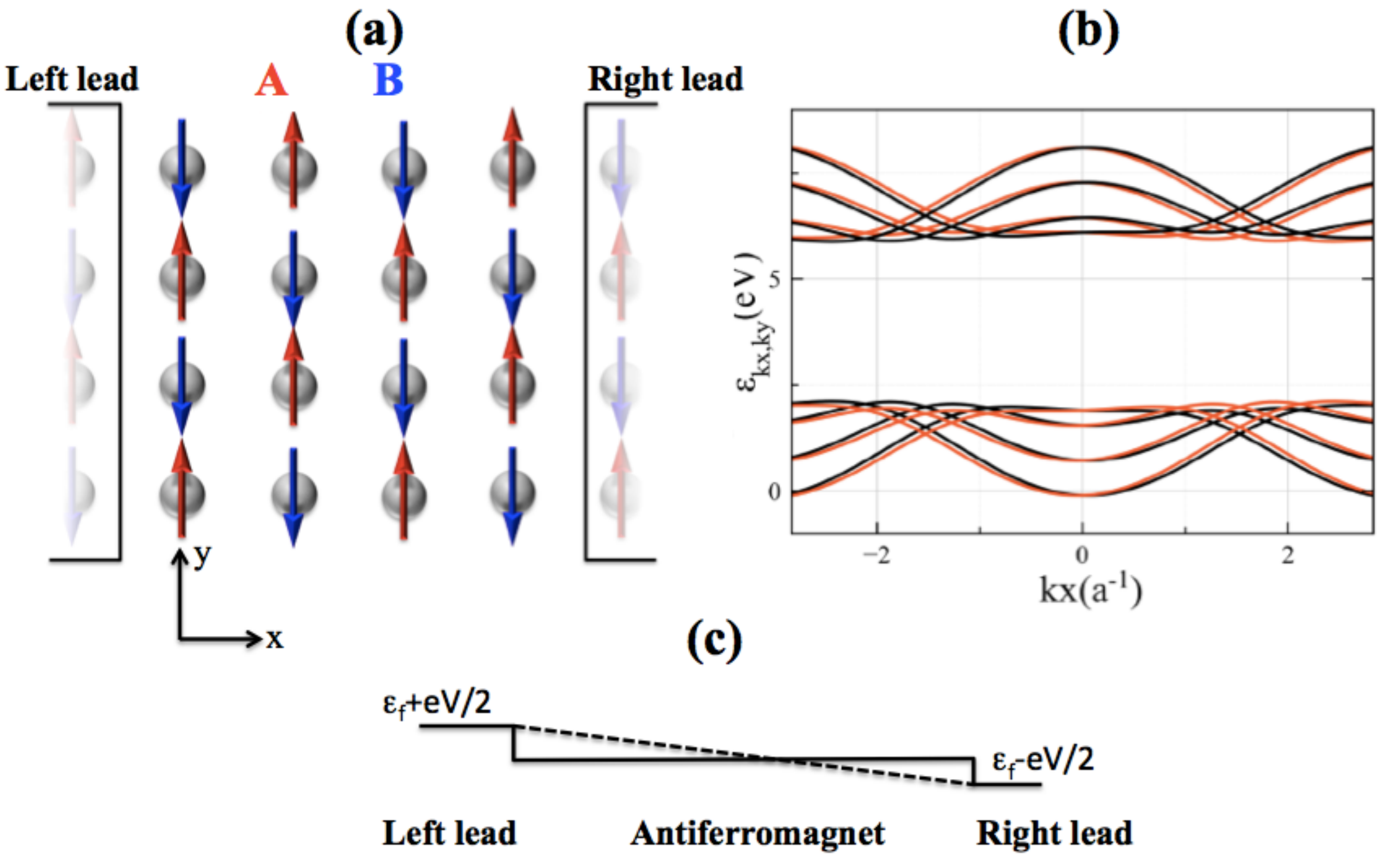}
  \caption{\small{(Color online) (a) Representation of the two dimensional antiferromagnetic system. The system is composed of central antiferromagnet with Rashba spin orbit interaction, connected to left and right leads. (b) The first four electronic bands of the antiferromagnet in the presence of Rashba spin orbit coupling with $\alpha=t_{\rm so}a = 0.04$ eV$\cdot$a and $J_{\rm ex}=2$. (c) Schematic of the potential profile used to extract intraband and interband contributions.} \label{fig:introductory}}
\end{figure}

Our system is composed of a two dimensional layer in the (x, y) plane connected to a left and right leads, as depicted on Fig. \ref{fig:introductory}(a). Similarly to the previous section, we consider a G-type antiferromagnet with Rashba spin-orbit coupling. The Hamiltonian reads
\begin{eqnarray}\label{eq:htb}
{\hat H_s}= &&\sum_{i,j}\{{\hat c}_{i,j}^+ (\epsilon_{i,j}+(-1)^{i+j}J_{\rm ex}{\bf n}\cdot{\hat{\bm\sigma}}){\hat c}_{i,j}+h.c.\}\\
&&-\sum_{i,j}t({\hat c}_{i+1,j}^+ {\hat c}_{i,j}+{\hat c}_{i,j+1}^+ {\hat c}_{i,j}+h.c.) \nonumber\\
&&+it_{so}\sum_{i,j}(-{\hat c}_{i+1,j}^+ \hat{\sigma}_y{\hat c}_{i,j}+{\hat c}_{i,j+1}^+\hat{\sigma}_x {\hat c}_{i,j}+h.c.).\nonumber
\end{eqnarray}
The indices ($i,j$) represent the atomic positions in the (x, y) plane. The first two terms represent for the on-site energies ($\sim\epsilon_{i,j}$) and the exchange interaction ($\sim J_{\rm ex}$) between local moments and itinerant spins respectively. The third term is the hopping energy between nearest neighbors ($\sim t$) and the fourth term is the Rashba interaction $\sim t_{so}$. The Rashba energy is related to the Rashba parameter by $\alpha= t_{so}a$, where $a$ is the lattice parameter. In order to model disorder, the on-site energy $\epsilon_{i,j}$ can be varied from site to site by an amount proportional to the disorder strength of the system, $\epsilon_{i,j}\rightarrow{\bf \epsilon}_{i,j}+ \zeta_{i,j} \Gamma/2$, $\zeta_{i,j}$ being a random number between -1 and 1. Note that the disorder is spin-independent. ${\bf n}=(\cos\theta,\sin\theta,0)$ gives the direction of the N\'eel order parameter, $\theta$ being for the angle between the magnetic moment and the x axis. Finally, ${\hat c}_{i,j}^+$ is the creation operator of an electron on site $i,j$ such that ${\hat c}_{i,j}^+=(c_{i,j\uparrow}^+,c_{i,j\downarrow}^+)$, where $\uparrow,\downarrow$ refer to the spin projection along the quantization axis.\par

\subsection{Computing intraband and interband contributions}

To calculate the transport properties of this system, we use the non-equilibrium Green's function technique implemented on KWANT \cite{kwant}, following a procedure outlined in Ref. \onlinecite{saidaoui2014}. To promote charge transport through the system, a bias voltage $V$ is applied throughout the conductor such that $\mu_l-\mu_r=eV$ where $\mu_{l(r)}$ is the chemical potential of the left (right) reservoir. In this framework, all the information (charge/spin currents and densities) is contained in the lesser Green's function $\hat{G}^<$. The non-equilibrium spin density is computed by considering electrons coming from both left and right electrodes such that
\begin{eqnarray}\label{eq:negf}
{\bf s}=\int_{-\infty}^{+\infty}d\epsilon[{\bf s}_{lr}(\epsilon)f_l(\epsilon)+{\bf s}_{rl}(\epsilon)f_r(\epsilon)],
\end{eqnarray}
where ${\bf s}_{\alpha\beta}=(1/2\pi){\rm Tr}[\hat{\bm\sigma}\hat{G}^<_{\alpha\beta}]$ is the spin density of electrons originating from lead $\alpha$ and flowing toward lead $\beta$ and $f_\alpha(\epsilon)$ is the Fermi distribution in lead $\alpha$.\par

When an external electric field is applied on a material, it distorts both the electron (Fermi-Dirac) distribution and the electron wave function. Both effects give rise to non-equilibrium properties and can be considered as independent to the first order in electric field: the distortion of the Fermi distribution while keeping the wave functions unchanged results in {\em intraband} contributions, whereas the distortion of the wave function while keeping the Fermi distribution unmodified results in {\em interband} contributions. In translationally invariant systems and in the weak impurity limit, these contributions reduce to Eqs. (\ref{eq:intra}) and (\ref{eq:inter2}), respectively. However, when a conductor is attached to external leads, such as the one considered in this section (see Fig. \ref{fig:introductory}), the system lacks translational invariance and Kubo formula does not apply. Hence, the separation between intraband and interband contributions is rather subtle.\par

The intraband contributions, which usually constitute the largest contribution to non-equilibrium properties, can be readily computed in our real-space tight-binding model by simply assuming a (small) potential step between the leads and the conductor, as illustrated by the solid line in Fig. \ref{fig:introductory}(c). Since the potential in the conductor is kept constant, the wave functions of the conductor are not distorted by the electric field and only intraband contributions are taken into account. In the limit of small bias, the result is equivalent to considering only electrons flowing from left to right, and one recovers Landauer-Buttiker formula, ${\bf s}\approx {\bf s}_{lr}(\epsilon_{\rm F})eV$. \par

Nevertheless, since the potential profile is flat in the conduction region [solid line in Fig. \ref{fig:introductory}(c)], the wave functions in the conductor are left unaltered and this method does not compute the interband contributions. To do so, one needs to consider that the actual potential profile in the conductor. This can be done by self-consistently solving Schr\"odinger and Poisson equations, which is out of reach of our computational capabilities. Therefore, to account for the effect of the potential gradient in the conductor, we assume a linear potential profile [dashed line in Fig. \ref{fig:introductory}(c)] that connects the electrochemical potentials of the left and right leads. Since our calculations are restrained to the linear response of the conductor, we expect the deviations from the "real potential" case to be reduced in the limit of small bias voltages.\par

As a result, Eq. (\ref{eq:negf}) can be parsed into three terms ${\bf s}={\bf s}_{\rm eq}+{\bf s}_{\rm intra}+{\bf s}_{\rm inter}$, such that
\begin{eqnarray}\label{eq:seq}
{\bf s}_{\rm eq}&=&\int_{-\infty}^{\epsilon_{\rm F}}d\epsilon[{\bf s}^0_{lr}(\epsilon)+{\bf s}^0_{rl}(\epsilon)],\\\label{eq:sintra}
{\bf s}_{\rm intra}&=&[{\bf s}^0_{lr}(\epsilon_{\rm F})-{\bf s}^0_{rl}(\epsilon_{\rm F})](eV/2)\\\label{eq:sinter}
{\bf s}_{\rm inter}&=&\int_{-\infty}^{\epsilon_{\rm F}}d\epsilon[\partial_{eV}{\bf s}_{lr}(\epsilon)+\partial_{eV}{\bf s}_{rl}(\epsilon)](eV/2)
\end{eqnarray}
where ${\bf s}_{\alpha\beta}\approx{\bf s}_{\alpha\beta}^0+\partial_{eV}{\bf s}_{\alpha\beta}eV/2$ and $f_{\alpha}(\epsilon)=f^0(\epsilon)\pm\delta(\epsilon-\epsilon_{\rm F})eV/2$. The first term, ${\bf s}_{\rm eq}$, is calculated in the absence of bias voltage and produces the equilibrium magnetic anisotropy. The second term, ${\bf s}_{\rm intra}$, is driven by electron distribution imbalance between the left and right electrodes and produces the intraband contribution and finally, the third term, ${\bf s}_{\rm inter}$, involves the distortion of the wave function by the local gradient of electric potential and produces the interband contributions.\par

Finally, in order to give a transparent account of the magnitude of spin densities and torque on each sublattice, we define the uniform and staggered spin densities as ${\bf s}_{\rm u}={\bf s}_A+{\bf s}_B$ and ${\bf s}_{\rm st}={\bf s}_A-{\bf s}_B$. According to the analysis of the antiferromagnetic dynamics \cite{gomonay, zelezny2014,zelezny2016, Jungwirth2016}, a uniform spin density acts as a magnetic field on the antiferromagnet and therefore, only its time derivative, $\partial_t{\bf s}_{\rm u}$, can exert a torque on the antiferromagnetic order parameter. In contrast, the staggered spin density ${\bf s}_{\rm st}$ exerts an efficient torque on the antiferromagnetic order parameter that enables its electrical manipulation. Of course, both the uniform and staggered spin densities possess in-plane ($\sim{\bf y}$) and out-of-plane ($\sim{\bf n}\times{\bf y}$) components. The unit vector ${\bf y}\propto{\bf z}\times e{\bf E}$ is defined by the symmetry of Rashba spin-orbit interaction.

\subsection{Interband vs Intraband contributions\label{s:extint}}

\begin{figure}[h!]
  \includegraphics[width=8.5cm]{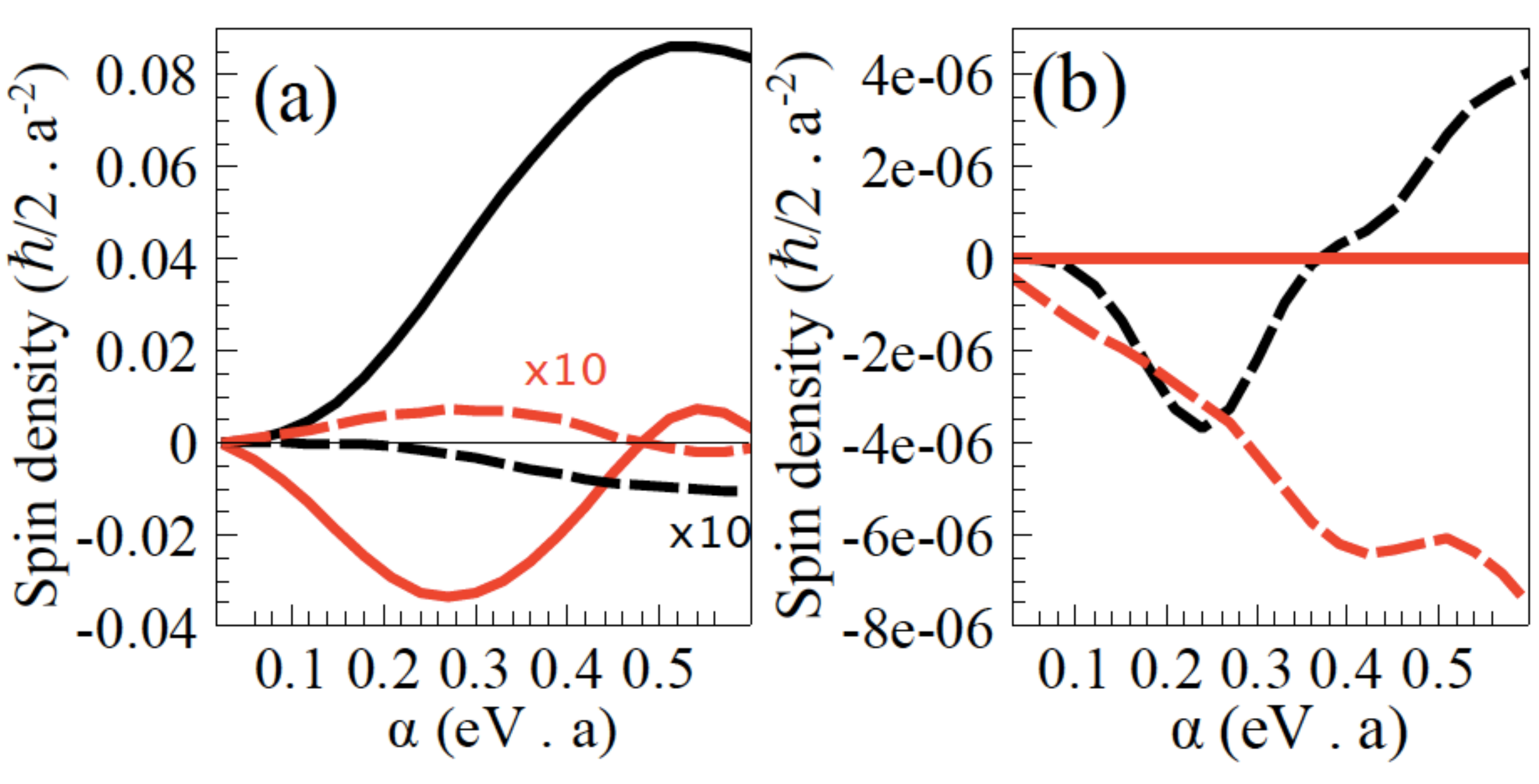}
  \caption{\small{(Color online) Dependence of the (a) uniform and (b) staggered spin densities as a function of the spin-orbit parameter $\alpha$. The black solid (dashed) line refers to in-plane ($\sim {\bf y}$) intraband (interband) component, while the red solid (dashed) line refers to out-of-plane ($\sim{\bf n}\times{\bf y}$) intraband (interband) component. The parameters are $J_{\rm ex}=2$ eV and $\epsilon_{\rm F}=0.1$ eV.}\label{fig:ev_alpha}}
\end{figure}

The spin density calculated using Eqs. \eqref{eq:sintra} and \eqref{eq:sinter} as a function of the Rashba spin-orbit coupling parameter are shown in Fig. \ref{fig:ev_alpha}. In these calculations, the order parameter is aligned along ${\bf n}={\bf x}$ and the parameters are given in the figure caption. The uniform spin density is reported on Fig. \ref{fig:ev_alpha}(a), while the staggered spin density is reported on Fig. \ref{fig:ev_alpha}(b). The solid (dashed) lines denote the intraband (interband) contributions, while the black (red) color refers to ${\bf y}$ and ${\bf z}$ components.

The non-equilibrium spin density is dominated by the uniform $s_y$ component, which possesses a large intraband contribution and small interband contribution [black lines in Fig. \ref{fig:ev_alpha}(a)]. This result is consistent with our analytical model and with previous works on inverse spin galvanic effect of magnetic Rashba two dimensional electron gas \cite{manchon2008,bijl2012,Pesin2012,wangmanchon2012,Li2015}. We also obtain a uniform $s_z$ component [red lines in Fig. \ref{fig:ev_alpha}(a)], which is absent from the Kubo formula calculations (Section \ref{s:kubo} and Refs. \cite{zelezny2014,zelezny2016}). This component is attributed to the electron reflections occurring at the interface between the conductor and the leads.\par

The staggered spin density is reported in Fig. \ref{fig:ev_alpha}(b). Interestingly, these spin densities are three to four orders of magnitude smaller than the uniform spin density (but still well above our numerical accuracy) and only arise from interband contributions. Intraband contributions do not produce staggered spin density, consistently with the Kubo formula calculations (Section \ref{s:kubo} and Refs. \cite{zelezny2014,zelezny2016}). Notice that the staggered spin density is dominated by $s_z$ component, while the large staggered $s_y$ component is again attributed to electron reflection at the interfaces with the leads.\par

One comment is in order at this stage. The intraband and interband contributions that we calculated in the previous section in the weak disorder limit produce so-called {\em extrinsic} and {\em intrinsic} contributions to the spin-orbit torque. In other words, the intraband contribution arising from the Fermi surface is proportional to the impurity concentration ($\sim 1/\Gamma$) while the interband contribution arising from the Fermi sea is independent of it when $\Gamma\rightarrow0$. Therefore, one could argue that in a ballistic calculation such as the one performed in the present section, the intraband contribution [Eq. (\ref{eq:sintra}), solid lines in Fig. \ref{fig:ev_alpha}(a)] should only produce a uniform $s_y$ component, while the interband contribution [Eq. (\ref{eq:sinter}), dashed lines in Fig. \ref{fig:ev_alpha}(a) and (b)] should only produce a staggered $s_z$ component. This reasoning is clearly inadequate as shown by the results obtained in Fig. \ref{fig:ev_alpha}. Indeed, we are performing numerical simulations on a finite size system in which the quantum confinement entirely controls the transport properties, in sharp contrast with disorder-dominated bulk transport. Therefore, it is not surprising that our calculations produce both uniform and staggered $s_y$ and $s_z$ spin densities. Actually, as will be discussed in the Section \ref{s:dis}, the uniform $s_z$ component is much less robust against disorder than the uniform $s_y$ component, which reconciles the numerical simulation in finite size systems with the bulk calculations performed in Section \ref{s:kubo} and Ref. \onlinecite{zelezny2014,zelezny2016}.

\subsection{Impact of disorder on the antiferromagnetic spin-orbit torque\label{s:dis}}

Let us now investigate the impact of disorder on the spin-orbit torque, ${\bm\tau}=J_{\rm ex}{\bf n}\times{\bf s}$. To do so, the spin-orbit torque is averaged over 10$^5$ disorder configurations. Notice that testing the influence of the disorder on the {\em interband} contributions necessitates both energy integration and disorder configurational average, which is highly computationally demanding and was out of our reach. Therefore, the results discussed in this section apply to the spin-orbit torque arising from the uniform  {\em intraband} spin density only.

\subsubsection{Robustness of the antiferromagnetic spin-orbit torque}

\begin{figure}[h!]
  \centering
  \includegraphics[width=8cm]{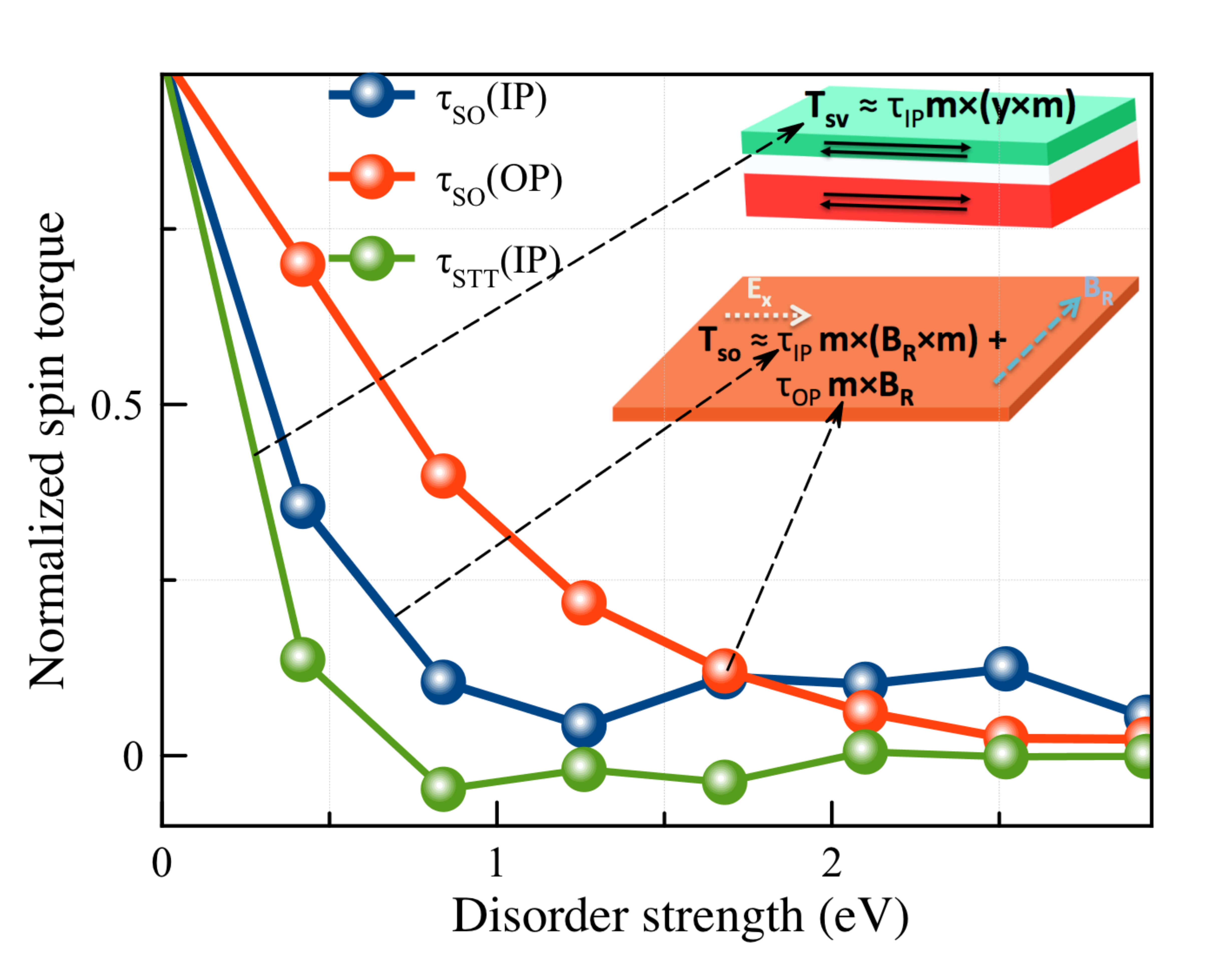}
  \caption{\small{(Color online) Dependence of the IP (blue symbols) and OP (red symbols) spin-orbit torque components as a function of the disorder strength for $J_{\rm ex}=2$ eV, $\epsilon_{\rm F}=-0.3$ eV and $\alpha= 0.3$ eV$\cdot$a. The IP spin torque calculated on an antiferromagnetic spin-valve for $J_{\rm ex}=2$ eV, $\epsilon_{\rm F}=6.4$ eV is also reported (green symbols). The torques are normalized to their value in the absence of disorder.}\label{fig:fig4}}
\end{figure}

In order to test the robustness of the antiferromagnetic spin-orbit torque against disorder, we calculated the intraband contributions to the spin-orbit torque as a function of the disorder strength $\Gamma$, as reported in Fig. \ref{fig:fig4}. The uniform $s_y$ component produces an out-of-plane (OP) torque, while the uniform $s_z$ component produces an in-plane (IP) torque. From Fig. \ref{fig:fig4}, it clearly appears that both IP and OP spin-orbit torques are reduced by disorder-induced scattering. However, the OP component (red symbols) is much more robust than the IP one (blue symbols). Therefore, in the diffusive regime in bulk conductors, one can reasonably anticipate that intraband contributions mostly produce an $s_y$ spin density component consistently with the results obtained in Section \ref{s:kubo}.\par

Let us now compare these results with the spin transfer torque in antiferromagnetic spin-valves. The metallic antiferromagnetic spin-valve is similar as Ref. \onlinecite{saidaoui2014}, for comparison. The results obtained for the largest spin torque component (staggered IP torque \cite{saidaoui2014}) are reported in Figure \ref{fig:fig4} (green symbols). The spin-orbit torque is in general much more robust against disorder than the spin transfer torque in antiferromagnetic spin-valves. Indeed, as mentioned in the introduction, spin transfer torque in spin-valves require the coherent transmission of the spin density throughout the device, while spin-orbit torque is a local torque and is therefore much less sensitive to momentum scattering.

\begin{figure}[h!]
  \centering
  \includegraphics[width=8cm]{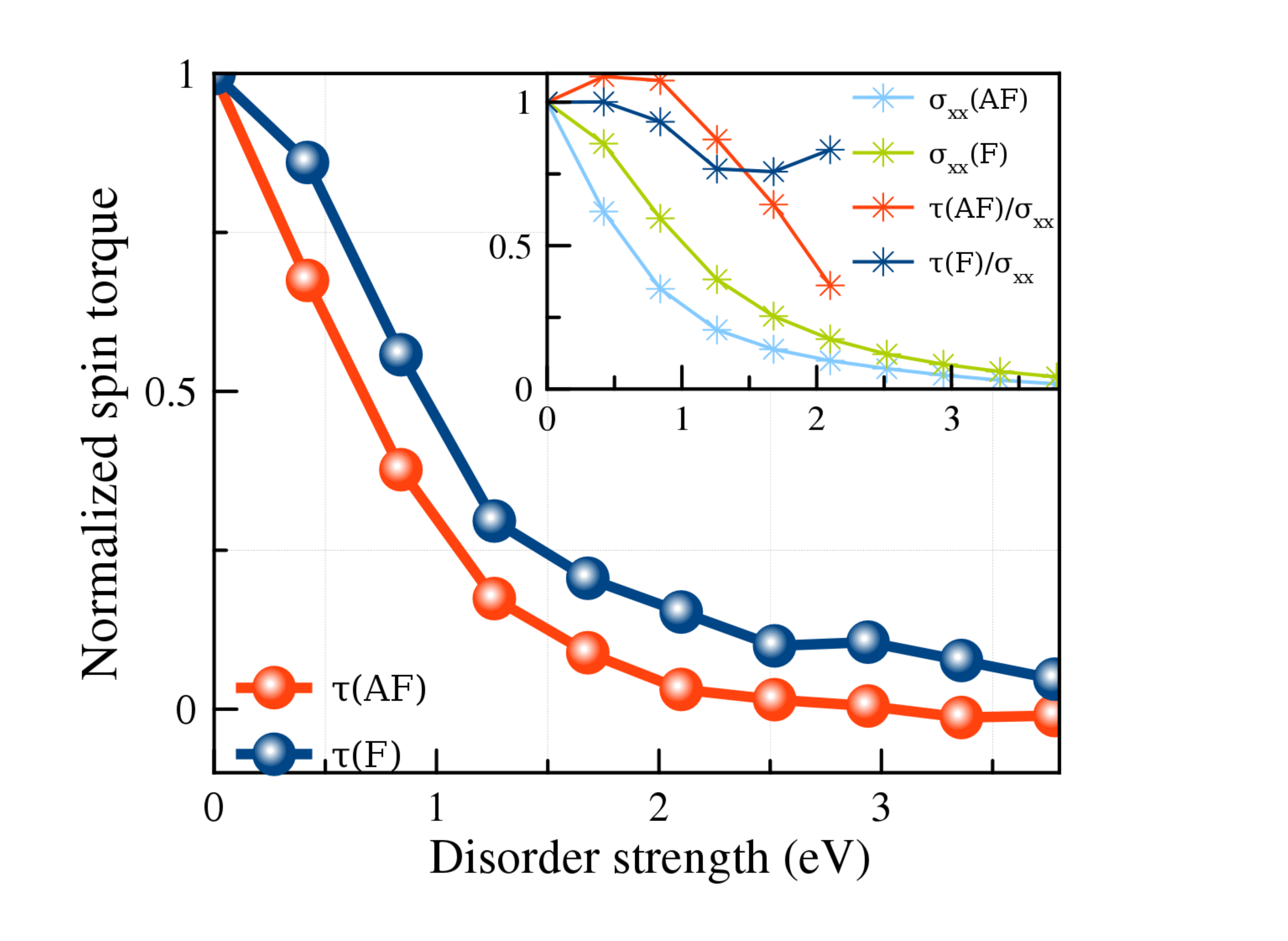}
  \caption{\small{(Color online) Dependence of the OP spin-orbit torque component as a function of the disorder strength in a ferromagnetic (blue symbols) and antiferromagnetic (red symbols) two dimensional electron gas. The torques are normalized to their value in the absence of disorder. The inset shows the corresponding normalized conductance and torque efficiency as a function of disorder strength. Blue (red) symbols correspond to the torque efficiency in ferromagnetic (antiferromagnetic) two dimensional, light green and light blue symbols stand for the corresponding conductances. The parameters are $J_{\rm ex}=2$ eV, $\epsilon_{\rm F}=-0.3$ eV and $\alpha= 0.3$ eV$\cdot$a} \label{fig:Torque-AF-F}}
\end{figure}

To complete this study, we have finally compare the impact of disorder on OP spin-orbit torque in a ferromagnetic Rashba gas in Fig. \ref{fig:Torque-AF-F}. Both torques are fairly robust against disorder. The ferromagnetic spin-orbit torque conserves about half its magnitude for $\Gamma=1$ eV, which corresponds to a mean free path of $20$ atomic sites while the antiferromagnetic spin-orbit torque is reduced to 25\% of its ballistic strength. Notice though that in the case of the ferromagnetic Rashba gas, a significant part of the torque reduction can be attributed to the increased resistivity as shown in the inset of Fig. \ref{fig:Torque-AF-F}: the torque efficiency, defined as the ratio between the torque and the conductivity, is less sensitive to disorder in the ferromagnetic case than in the antiferromagnetic one.

\subsubsection{General dependences of the spin-orbit torque}

\begin{figure}[h!]
  \centering
  \includegraphics[width=8cm]{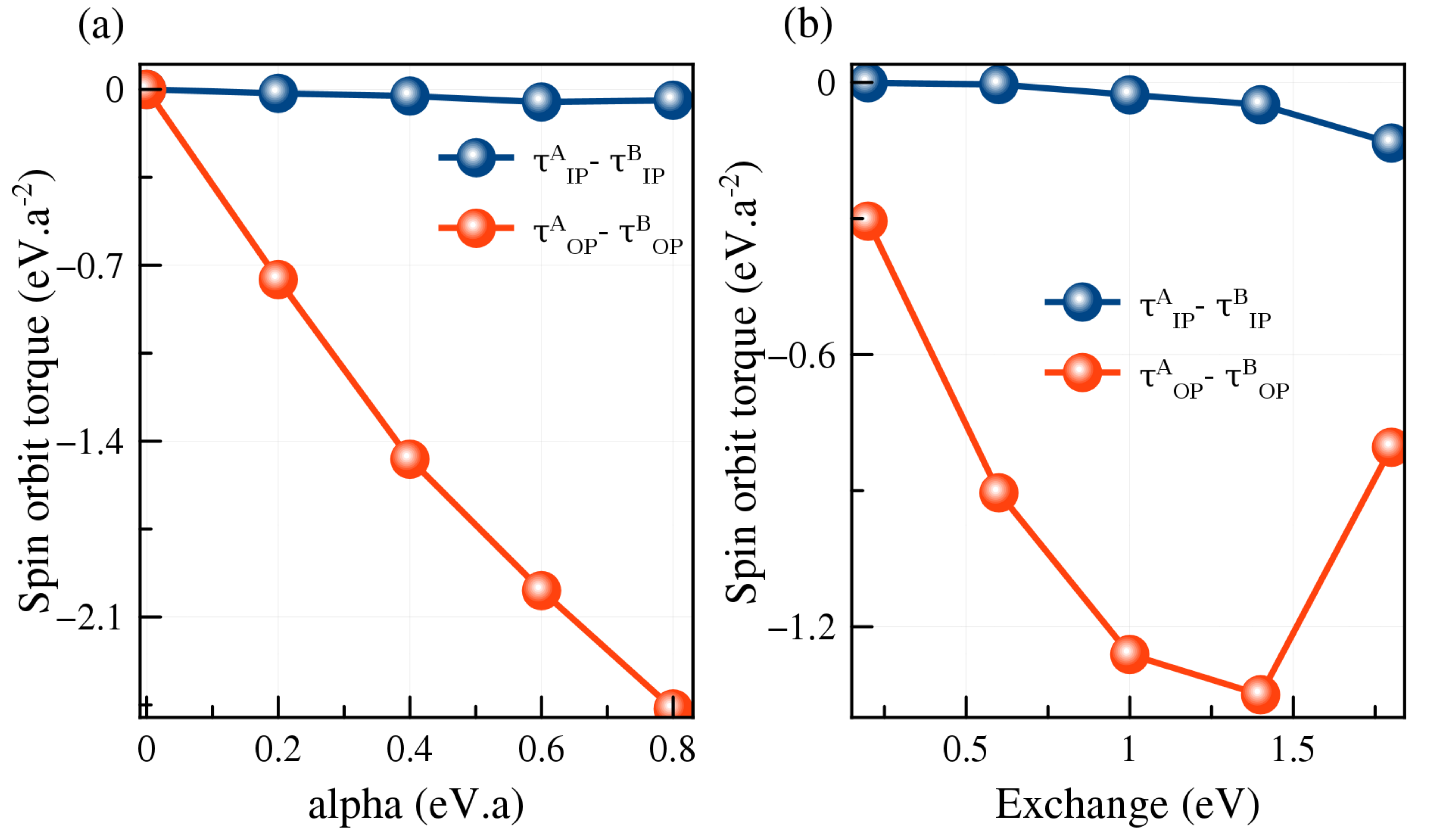}
  \caption{\small{(Color online) (a) Dependence of the IP and OP spin-orbit torque components exerted on the antiferromagnet on the spin-orbit strength $\alpha$ for $J_{\rm ex}=2$ eV, $\epsilon_{\rm F}=-0.3$ eV. (b) Dependence of the IP and OP spin-orbit torque components exerted on the antiferromagnet on the exchange strength $J_{\rm ex}$ for $\epsilon_{\rm F}=1$ eV and $\alpha= 0.3$ eV$\cdot$a. The dependence is calculated in the presence of disorder with $\Gamma = 1$ eV.}  \label{fig:torque_alpha}}
\end{figure}

Let us now turn our attention towards the dependence of the torque as a function of material parameters in the disordered regime ($\Gamma=1$ eV, which corresponds to mean free path of about 20 atomic sites). In the following we choose ${\bf n}={\bf x}$, i.e. the magnetization lies along the electric field direction. Fig. \ref{fig:torque_alpha} reports the variation of the IP and OP components of the staggered torque (stemming from the uniform spin density) as function of the spin-orbit strength $\alpha$ [see Fig. \ref{fig:torque_alpha}(a)] and exchange $J_{\rm ex}$ [see Fig. \ref{fig:torque_alpha}(b)].

As expected from Fig. \ref{fig:fig4}, the IP component vanishes in the presence of disorder and remains almost zero for the parameter range considered. The OP component displays the regular behavior expected for the intraband Rashba torque due to inverse spin galvanic effect \cite{manchon2008,Li2015}. It increases with $\alpha$, and displays a non-linear dependence as a function of the exchange. It first increases with exchange from $0.2$ to $\sim 1.25$ eV, and then decreases beyond this point. This decrease is similar to the one obtained in Ref. \onlinecite{Li2015} in the case of (Ga,Mn)As. Notice though that in antiferromagnetic Rashba gas, increasing the exchange also leads to the enhancement of the band gap and the reduction of the conductivity. Such reduction is also expected to impact the magnitude of the torque in the strong exchange limit, although it is difficult to quantify its influence.

\section{Conclusion\label{s:conclusion}}
In this work, we investigated the nature of current-driven spin-orbit torques in an antiferromagnetic electron gas with Rashba spin-orbit coupling, using both analytical and numerical methods. In agreement with \v Zelezn\'y et al. \cite{zelezny2014,zelezny2016}, we found that the intraband contribution produces an out-of-plane torque, which is inefficient to electrically manipulate the antiferromagnetic order parameter, while the interband contribution produces a {\em staggered} spin density that enables the order parameter manipulation. Interestingly, the torques we obtain are much more robust against disorder than the one previously obtained in metallic antiferromagnetic spin-valves\cite{saidaoui2014}, indicating that interfacial spin-orbit coupling is a viable route towards the electrical manipulation of antiferromagnets. 

\acknowledgments
This work was supported by the King Abdullah University of Science and Technology (KAUST) through the Office of
Sponsored Research (OSR) [Grant Number OSR-2015-CRG4-2626]. A.M. acknowledges inspiring discussions with X. Waintal, T. Jungwirth, J. Sinova, H. Gomonay and J. Zelezny.


\begin{thebibliography}{99}
\bibitem{comment} The factors ($1/8\pi$) in Eq. (\ref{eq: Sintaraf}) and ($1/2\pi$) in Eq. (\ref{eq: Sintraf}) since the antiferromagnetic has a diatomic unit cell. A consequence of this is that the energy broadening is $\Gamma=\hbar/(4\tau)$ in antiferromagnets while is it $\Gamma=\hbar/(2\tau)$ in ferromagnets, $\tau$ being the momentum relaxation time of a given quantum state. Furthermore, the spin density given in Eq. (\ref{eq: Sintaraf}) is for one site of the unit cell only. Therefore, per unit cell, Eqs. (\ref{eq: Sintaraf}) and (\ref{eq: Sintraf}) are equivalent.
\end{thebibliography}
\end{document}